\documentclass[twocolumn,showpacs,preprintnumbers,prl,amsmath,amssymb]{revtex4}

\usepackage[pdftex]{graphicx}
\usepackage{dcolumn}
\usepackage{bm}
\begin{document}

\title{Magnetoelastic coupling and  charge correlation lengths in a twin domain of Ba(Fe$_{1-x}$Co$_{x}$)$_{2}$As$_{2}$ ($x=0.047$): A high-resolution X-ray diffraction study}
\author{Qiang Zhang,$^{1}$}\email{qzhangemail@gmail.com} \author{Wenjie Wang,$^1$ Jong-Woo Kim,$^{2}$ Benjamin Hansen,$^{1}$ Ni Ni,$^{1}$ Sergey  L. Bud'ko,$^{1}$ Paul C. Canfield,$^{1}$ Robert J. McQueeney,$^{1}$ and David Vaknin$^{1}$} \email{vaknin@ameslab.gov}

\affiliation{$^1$Ames Laboratory, and Department of Physics and Astronomy, Iowa State University, Ames, Iowa 50011, USA \\
$^2$Advanced Photon Source, Argonne National Laboratory, Argonne, Illinois 60439, USA }

\date{\today}

\begin{abstract}
     The interplay between structure, magnetism and superconductivity in single crystal Ba(Fe$_{1-x}$Co$_{x}$)$_{2}$As$_{2}$ (x=0.047) has been studied using high-resolution X-ray 
diffraction by monitoring charge Bragg reflections in each twin domain separately. The emergence of the superconducting state is correlated with the suppression of the orthorhombic 
distortion around \emph{T}$_\texttt{C}$, exhibiting competition between orthorhombicity and superconductivity.  Above \emph{T}$_\texttt{S}$, the in-plane  charge correlation length increases
with the decrease of temperature, possibly induced by nematic fluctuations in the paramagnetic tetragonal phase. Upon cooling, anomalies in the in-plane charge correlation lengths along $a$ ($\xi_{a}$)
and $b$ axes ($\xi_{b}$) are observed at \emph{T}$_\texttt{S}$ and also at \emph{T}$_\texttt{N}$ indicative of strong magnetoelastic coupling. The in-plane charge 
 correlation lengths are found to exhibit anisotropic behavior along and perpendicular to the in-plane component of stripe-type AFM wave vector (101)$_{\rm O}$ below around \emph{T}$_\texttt{N}$. The temperature dependence of the out-of-plane  charge correlation length shows a single anomaly at  \emph{T}$_\texttt{N}$, reflecting the connection between Fe-As distance and Fe local moment. 
 The origin of the anisotropic in-plane charge correlation lengths $\xi_{a}$ and $\xi_{b}$ is discussed on the basis of the antiphase magnetic domains and their dynamic fluctuations.     
\end{abstract}

 \maketitle
\section{Introduction}
In the recently discovered iron-based superconductors \cite{Kamihara2008,Aswathy2010}, the superconducting  temperatures are found to be in close proximity to an antiferromagnetic (AFM)
 and a tetragonal-orthorhombic (T-O) structural transitions. It turns out that the suppression of both the AFM and the T-O structural transitions, by doping or by pressure, eventually induces superconductivity. These phenomena beg
the question about the role of spin and lattice degrees-of-freedom in the emergence of superconductivity\cite{Aswathy2010,Fernandes2010}. Several theoretical descriptions have been proposed
 to interpret the relationship between the structural and AFM transitions on the basis of orbital ordering or by introducing an intermediate spin-nematic phase resulting from an
effective $J_1-J_2$ local-spin model or from an itinerant model both with equivalent consequences.\cite{Fernandes2010,Barzykin2009,Cano2010,Fernandes2012}
Although different in details, these descriptions emphasize the importance of the magnetoelastic coupling in driving the two transitions simultaneously or separately. Cano\emph{et al.} \cite{Cano2010}
 studied the interplay between the elastic and spin degrees-of-freedom in iron pnictide superconductors using a Ginzburg-Landau approach, indicating that the magnetoelastic coupling can bring about
the particular features of the structural and magnetic transitions in these systems including the emergence of the collinear stripe-type AFM ordering. Recently, a microscopic 
study \cite{Paul2011} of a simple symmetry-allowed model Hamiltonian demonstrated that due to the effect of magnetoelastic coupling, 
the considerable orthorhombic elastic softening is caused by critical spin fluctuations present in the system before magnetic order occurs. This may explain why the AFM transition is often preceded by the T-O 
structural transition. It should be pointed out that this picture is similar to the nematic phase model\cite{Fernandes2010,Fernandes2012}. To date,
there are very few experimental reports on the magnetoelastic effect in iron pnictides and such reports investigated the role that magnetoelastic effect plays in the structural and magnetic transitions under the application of external driving forces. 
Magnetoelastic effects have been demonstrated by applying pressure to CaFe$_2$As$_2$ and inducing an O-T  and AFM-to-nonmagnetic transitions\cite{Kreyssig2008}, or by applying shear stress to BaFe$_{2}$(As$_{1-x}$P$_{x}$)$_{2}$ 
that shifts the magnetic transition and superconducting critical temperatures significantly.\cite{Kuo2012} Therefore, it is of interest to study the possibly intrinsic magnetoelastic effect in iron-based superconductors, without introducing any external driving force.
          
 The Co-doped BaFe$_{2}$As$_{2}$ system exhibits a rich phase diagram with a complex interplay between the structural, magnetic, and superconducting phases.\cite{Nandi2010,Pratt2009,Ni2008,Chu2009}
 In the parent BaFe$_{2}$As$_{2}$ compound, the AFM ordering transition at \emph{T}$_\texttt{N}$ coincides with a T-O structural transition at \emph{T}$_\texttt{S}$. Upon doping both transitions gradually
separate, such that \emph{T}$_\texttt{S}$ $>$ \emph{T}$_\texttt{N}$, accompanied with the appearance of superconductivity above $x=0.03$ in Ba(Fe$_{1-x}$Co$_x$)$_2$As$_{2}$. It is interesting to point out that the orthorhombic distortion $\delta$ in 
Ba(Fe$_{1-x}$Co$_x$)$_2$As$_{2}$ with lower Co content ($x=0.018$) shows one clear anomaly at the magnetic transition temperature \emph{T}$_\texttt{N}$, but it is absent at \emph{T}$_\texttt{N}$ 
for $x=0.047$ superconductor with intermediate Co content. For $x$  higher than $\sim$0.066, both the magnetic and structural 
transitions are completely suppressed and superconducting transition is the only transition observed. Only in an intermediate composition region of $0.03\lesssim x\lesssim 0.066$, does Ba(Fe$_{1-x}$Co$_{x}$)$_{2}$As$_{2}$ exhibit a coexistence of
 superconductivity, O-structure and AFM phases providing potential candidates to investigate the effects of the magnetoelastic coupling. However,the tendency of these crystals to form twinned orthorhombic domains
 has hampered definitive determination of inherent features of the intermediate phase between \emph{T}$_\texttt{S}$ and \emph{T}$_\texttt{N}$ where the presumed nematic phase exists. Therefore, there have been extensive efforts to de-twin these crystals \cite{Chu2010} to establish the underlying
electronic, structural, and magnetic anisotropies that characterize this intermediate phase. Motivated by these issues, we set out to investigate anisotropic features of the crystal structure and domain formation over a wide range of temperatures using high resolution x-ray diffraction methods 
that reveals the intrinsic magnetoelastic coupling in the Ba(Fe$_{1-x}$Co$_{x}$)$_{2}$As$_{2}$ ($x=0.047$) superconductor.  
The high resolution allows us to separately monitor Bragg reflections of different orthorhombic twin domains and study their temperature evolution, as has been done recently on CeFeAsO\cite{Li2011}.

\begin{figure}\centering \includegraphics [width = 1\linewidth] {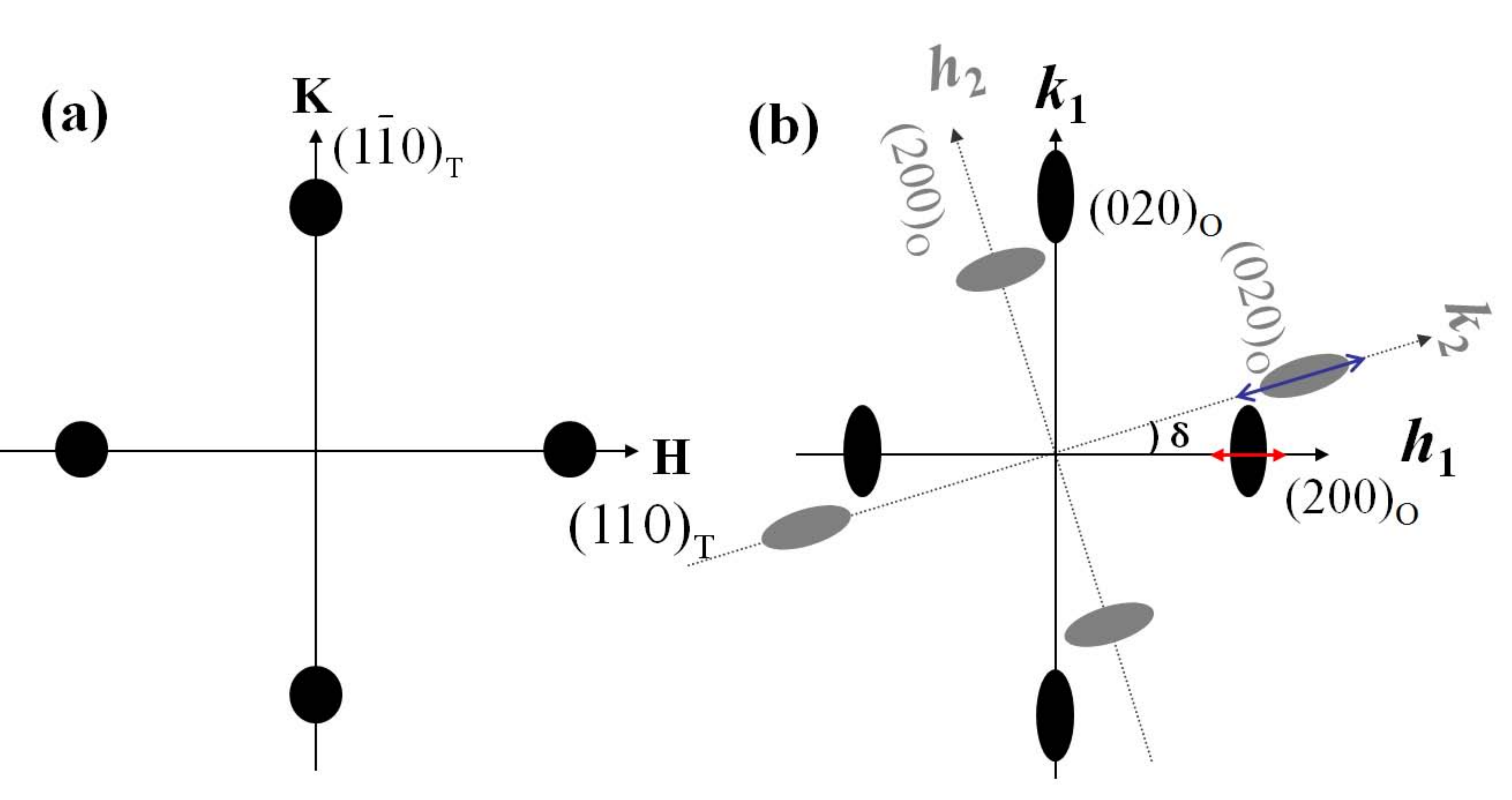}
\caption{Schematic illustration in reciprocal space of (a) untwinned crystal in the (\emph{H}\emph{K}0)$_{\rm T}$ plane at the high-temperature tetragonal phase and (b) at the orthorhombic phase with uniformly rotated and separated anisotropic twin-domains in (\emph{h}\emph{k}0)$_{\rm O}$ plane.  The arrows show typical scans performed at Bragg reflections.}
\label{fig1}
\end{figure}

\section{Experimental Details}
The Ba(Fe$_{1-x}$Co$_{x}$)$_{2}$As$_{2}$  ($x=0.047$) crystal was grown using a self-flux solution method as described previously\cite{Ni2008,Kim2011}. The crystal has been characterized by 
X-ray diffraction, neutron scattering, magnetization, resistivity and heat capacity, identifying reported three transitions, structural at \emph{T}$_\texttt{S}\simeq60$ K, magnetic  at \emph{T}$_\texttt{N}\simeq47$ K
 and SC at \emph{T}$_\texttt{C}\simeq17$ K.\cite{Pratt2009,Kim2011} A plate-like piece of crystal with its $c$-axis perpendicular to its surface was chosen
 for investigations by  high-resolution x-ray scattering technique using the six-circle diffractometer of the 6-ID-B beamline at the Advanced Photon Source (APS), Argonne National Laboratory (X-ray energy kept at 8 keV). The scattering geometry of our experiment is similar to that used previously by Li \emph{et al.}\cite{Li2011}. We use orthorhombic indices $(hkl)_{\rm O}$ at all temperatures so that  
the  tetragonal (\emph{HKL})$_\texttt{T}$ indices are provided in terms of the twin domains in the orthorhombic structure with the following conversion  $(H + K, H - K, L)_{\rm O}$ and $(H - K, H + K, L)_{\rm O}$. The crystal was mounted at the end of the cold-finger
 of a Displex cryogenic refrigerator with access to ($00l$)$_{\rm O}$ and high index ($hkl$)$_{\rm O}$ Bragg reflections.  Flux intensity on the sample was optimized to eliminate beam heating effects of the sample
 while maintaining a reasonable signal to noise ratio.  To achieve that, slit setups and attenuations yielding Bragg reflection intensities that scaled with beam attenuations were chosen. 
  In this regard, it should be noted that the low thermal conductivity in the SC state required a significant increase of beam attenuation to prevent sample-heating during measurements.  
 These considerations limited the choice of setups, i.e., resolution, but as discussed below, by adequate analytical tools we captured the intrinsic behavior (i.e.,  charge correlation lengths,
 in particular) of this system. As demonstrated on the CeFeAsO\cite{Li2011}, the twin domains are uniformly rotated and separated in reciprocal space
by a microscopic shear-angle enabling the characterization of 
each domain. Fig. 1 shows schematically a limited in-plane reciprocal zone with the (110)$_{\rm T}$ of the untwinned crystal (a) that transforms to the (200)$_{\rm O}$ and (020)$_{\rm O}$ in the orthorhombic 
symmetry notation (b). The misfit angle between the two domains is determined by the orthorhombic distortion ( $\delta\equiv\frac{a-b}{a+b}$).\cite{Li2011} The arrows in Fig.\ \ref{fig1} show typical scans performed at Bragg reflections.  In the present study,
 we monitored the (208)$_{\rm O}$/(028)$_{\rm O}$ systematically (corresponding to (118)$_\texttt{T}$ at high temperatures) and also the (008)$_{\rm O}$, as a function of temperature.  

\section{Results and Discussion}
Fig. 2(a) shows the evolution of the orthorhombicity as the temperature decreases with a sharp splitting of the (118)$_{\rm T}$ into the (208)$_{\rm O}$
and (028)$_{\rm O}$ at \emph{T}$_\texttt{S}= 60 $ K. With further decrease in temperature, the orthorhombicity increases without displaying
an anomaly at \emph{T}$_\texttt{N}= 47$ K (see the inset of Fig. 2(a)), which is consistent with the report by Kim \emph{et al.} \cite {Kim2011}. A slight decrease in the splitting is observed below the SC temperature \emph{T}$_\texttt{C}$, indicating that superconductivity and the orthorhombic distortion are coupled.\cite{Nandi2010}. Note that for higher Co substitution, the suppression of orthorhombic order parameters becomes larger. For example,
 the orthorhombic
distortion in $x=0.063$ is completely suppressed and the reentrant transition to tetragonal structure occurs below \emph{T}$_\texttt{C}$. When $x$ increases to 0.066, the orthorhombic distortion 
vanishes and no T-O transition is observed.\cite{Nandi2010} From Fig. 2 (a), we point out that  
the in-plane lattice constant shrinks linearly in the tetragonal phase of Ba(Fe$_{1-x}$Co$_{x}$)$_{2}$As$_{2}$ with decreasing temperature at a rate of $ 1.4\times10^{-5}$ {\AA}/K per unit cell (linear thermal expansion parameter $\alpha \sim 2.5\times10^{-6}$/K). 
 By contrast, the thermal expansion along the $c$-axis is weakly quadratic (Fig. 2(b)) with no abrupt anomaly at \emph{T}$_\texttt{S}$ and displays a deviation from the quadratic form near \emph{T}$_\texttt{N}$
 (a weak minimum is observed in \emph{c}-axis lattice parameter for (20\emph{l})$_{\rm O}$ reflection as shown in the inset of Fig.2(b)). The effect of magnetic transition at \emph{T}$_\texttt{N}$ on the lattice parameter $c$
implies a coupling between them, which will be discussed below. 

\begin{figure} \centering \includegraphics [width = 1\linewidth] {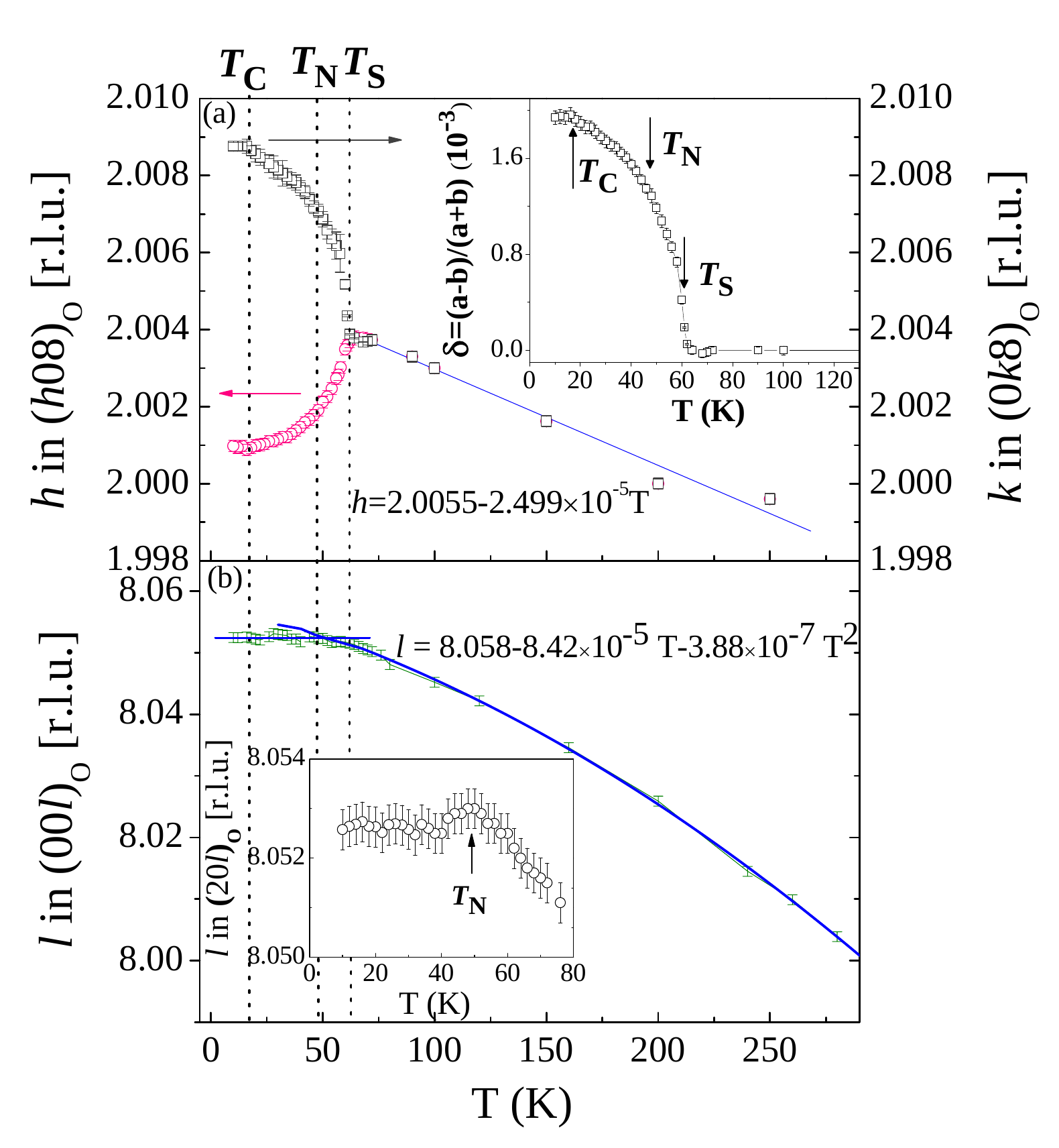}
\caption{(color online)  The temperature dependence of peak positions extracted from (a) (\emph{h}08)$_{\rm O}$  (\emph{h} domain, circles) and (0\emph{k}8)$_{\rm O}$ (\emph{k} domain,squares) scans,
 and (b) the \emph{l} scan for (00\emph{l})$_{\rm O}$ reflection.  The inset of (a) shows the orthorhombic distortion $\delta$ as a function of temperature. The inset of (b) shows the zoomed view on \emph{l} scan for (20\emph{l})$_{\rm O}$ reflection. Linear and quadritic fit to the data above \emph{T}$_\texttt{S}$ are included as solid lines. The dashed lines in (a) and (b) indicate the locations of structural, magnetic and superconducting transition temperatures  \emph{T}$_\texttt{S}$, \emph{T}$_\texttt{N}$, and \emph{T}$_\texttt{C}$.
}
\label{fig2} 
\end{figure}

        Fig. 3(a) shows representative in-plane scans along $h$ for the (208)$_{\rm O}$ Bragg reflection at 150 K (above \emph{T}$_\texttt{S}$) and 46 K (below \emph{T}$_\texttt{S}$).  
It is clear that the full width at half maximum (FWHM) of the Bragg peak is broader below \emph{T}$_\texttt{S}$. Variation in the line broadening can be
 due to changes in coherence length, domain size, mosaic distributions, and more likely a combination of all three. To obtain quantitative evaluation of peak line-widths, these and other
 scans were initially modeled as a Gaussian, a Lorentzian, their linear combination, or Pseudo-Voigt line-shapes, but none of these lineshapes yielded satisfactory agreement with the data. We therefore 
adopted a standard convolution method by systematically folding a Gaussian resolution function and a Lorentzian function that reflects an exponentially decaying charge (chemical) coherence length as follows,   

\begin{equation}
I(q)=\int_{-\infty}^{\infty} G(q')L(q-q')dq' \label{eq1}
\end{equation}

 where $G(x)=\frac{1}{\omega\sqrt{\pi\ln{2}}}\text{e}^{-(\ln{2})(x)^2/\omega^2}$ and $L(x)=\frac{C}{1+(x/\nu)^2}$, such that  $2\omega$ and $2\nu$ are the FWHM of the Gaussian and Lorentzian functions in reciprocal space, respectively. 
 Our resolution was high enough to resolve the twin domains separately, i.e., optimizing peak intensity of the (208)$_{\rm O}$ from one twin domain and that of the (028)$_{\rm O}$ of the other domain required sample rotation between the two peaks as has been done for
 CeFeAsO\cite{Li2011}. While the FWHM of the Lorentzian function ($2\nu$) represents the intrinsic width $\kappa$ of the sample, we should note that the Gaussian function in Eq.\ (\ref{eq1}) has two contributions: one from geometrical setup (i.e., incident and scattered beam divergence) and the other from the mosaic spread of the studied crystal 
(see a detailed discussion in resolution function in Ref.\ \cite{Cowley1987}). It is by now well established that the mosaic spread of typical pnictides undergoing shear induced O-T transition exhibit mosaic spread changes due to stresses during the transitions.  
We therefore attribute the temperature dependence of the Gaussian width as arising primarily from the variation in the mosaic distributions. Fig. 3(b) shows the FWHM of the Gaussian function for the
 (208)$_{\rm O}$ along $h$ as a function of temperature indicating anomalies that can be related to the stresses introduced by the structural and magnetic transitions in the system.  

\begin{figure} \centering \includegraphics [width = 1\linewidth] {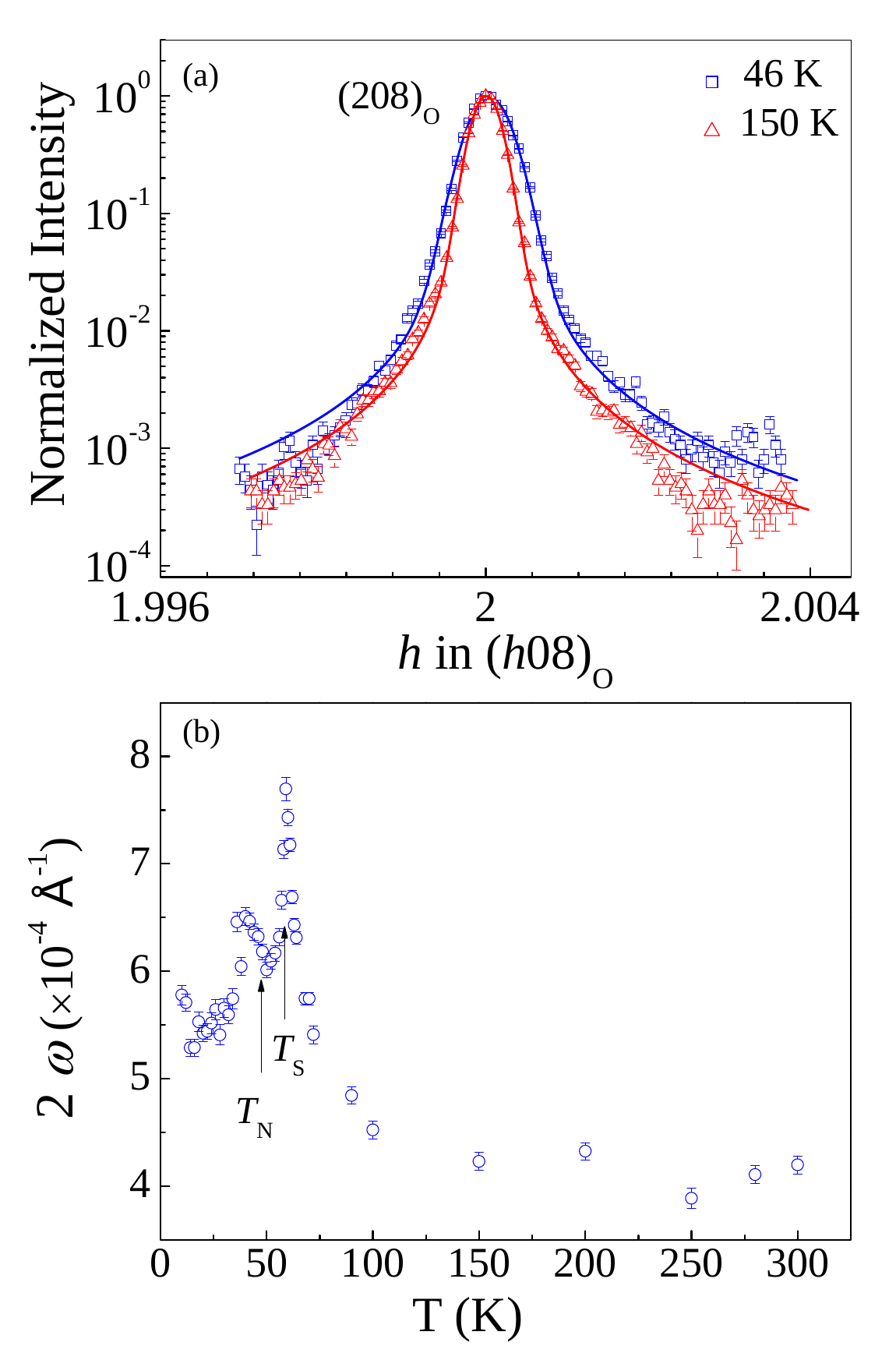}
\caption{(color online) (a) A representative longitudinal scan of \textbf{Q}=(\emph{h}08)$_{\rm O}$ at 150 K (above \emph{T}$_\texttt{S}$)) and 46 K (below \emph{T}$_\texttt{S}$)). 
The \textbf{Q} values and intensity were normalized for comparison. It can be seen that the peak widths broaden significantly below \emph{T}$_\texttt{S}$. (b) The temperature evolution of FWHM of the Gaussian function ($2\omega$) for the longitudinal scan of \emph{h} domain.}
\label{fig3} 
\end{figure}
           
       The temperature evolutions of the intrinsic width $\kappa$ ($=2\nu$) extracted from the Lorentzian functions of (208)$_{\rm O}$ and (028)$_{\rm O}$ Bragg peaks are shown in Fig.4(a). It is evident that
 anomalies are observed at both \emph{T}$_\texttt{S}$ and \emph{T}$_\texttt{N}$. Similar observation by using high-resolution x-ray diffraction was reported in different systems TbV$_{1-x}$As$_{x}$O$_{4}$ (x=0 and 1) \cite{Rule2008}, where the intrinsic
width of the Bragg peaks shows a clear peak at their \emph{T}$_\texttt{C}$ that is reminiscent of $\lambda$ anomalies in the heat capacity.
Note that the intrinsic width $\kappa$ in unit of $\AA^{-1}$, i.e., the FWHM of the Lorentzian profiles from X-ray scattering in reciprocal space corresponds to
the inverse charge correlation length $\xi$ \cite{Li2011,Rule2008,Feng2012,Lorenzo2008}:

\begin{equation}
\xi=1/\kappa
\end{equation}
       
Thus, the temperature dependence of the in-plane  charge correlation lengths along and normal to the in-plane component of
AFM propagation wavevector (101)$_{\rm O}$, i.e., longitudinal charge correlation length $\xi_{a}$ (along the AFM bond direction) and transverse charge correlation length $\xi_{b}$ (along the ferromagnetic bond direction), can be derived from the intrinsic widths of (208)$_{\rm O}$ and (028)$_{\rm O}$ Bragg peaks. As illustrated in Fig.4 (b), both $\xi_{a}$ and 
$\xi_{b}$ show two clear peaks at \emph{T}$_\texttt{S}$ and \emph{T}$_\texttt{N}$, respectively. With the decrease of temperature to \emph{T}$_\texttt{S}$, the in-plane  charge correlation lengths
increase gradually, followed by a rapid decrease below \emph{T}$_\texttt{S}$. When the temperature approaches \emph{T}$_\texttt{N}$, the in-plane  charge correlation lengths increase 
again. It is worth emphasizing that the Bragg reflections used to extract these data are not allowed by the symmetry of the  magnetic structure of the ordered iron moments and they are strictly
the result of charge (nuclei) ordering. Interestingly, the intrinsic width $\kappa$ and  charge correlation length obtained from the high-resolution X-ray data show a clear anomaly at the
 magnetic transition temperature, suggesting these charge Bragg reflections are sensitive to the spin-structure and fluctuations. This is presumably due to the strong magneto-elastic coupling that exerts secondary effects 
on charge correlations, domain formation and their shape. 

\begin{figure} \centering \includegraphics [width = 1\linewidth] {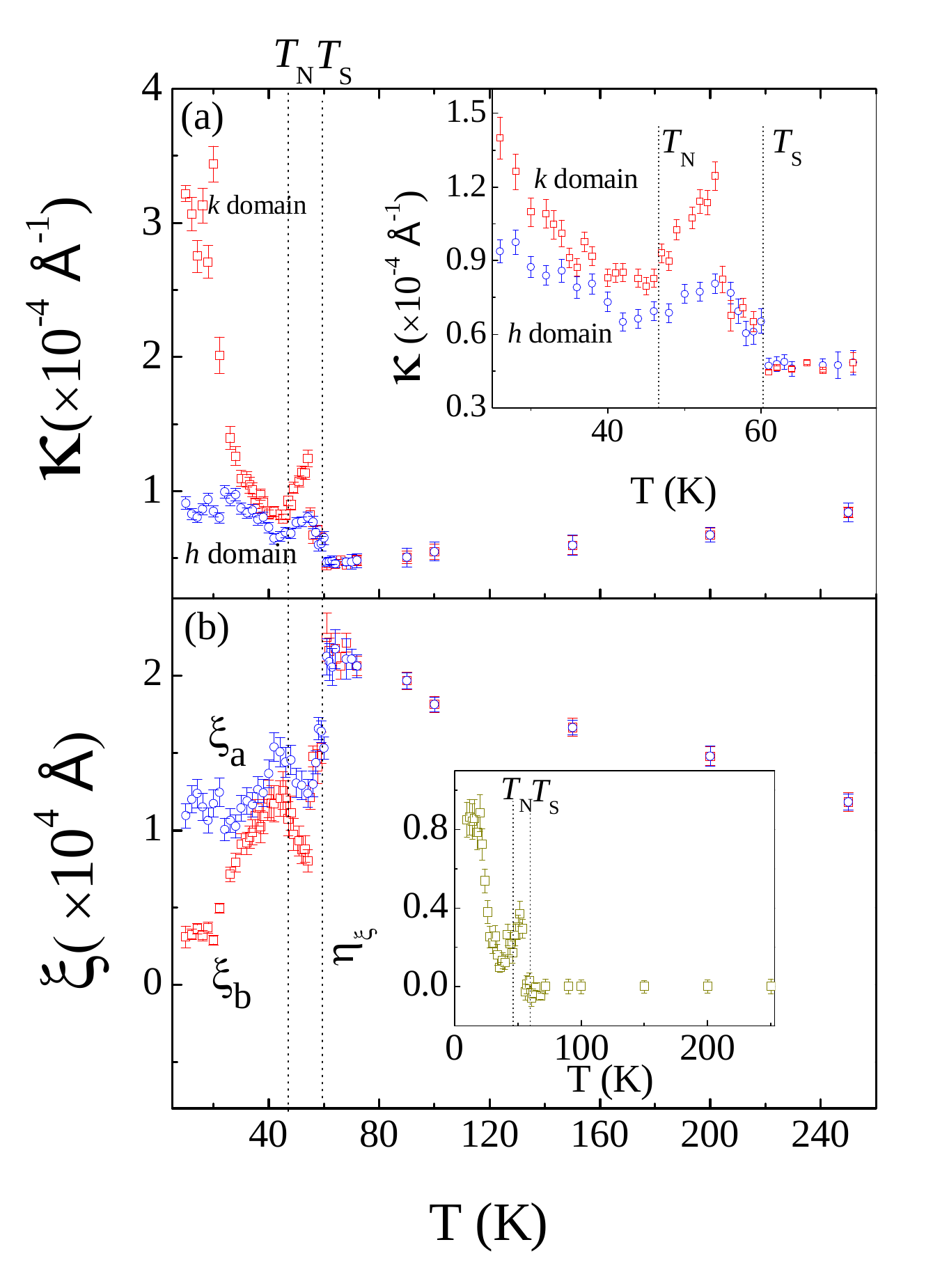}
\caption{(color online) The temperature dependence of (a) the intrinsic width $\kappa$ of the longitudinal scans of \emph{h} and \emph{k} domains obtained 
from the Lorentzian function and (b) the in-plane charge correlation lengths along $a$ axis ($\xi_{a}$) and $b$ axis ($\xi_{b}$). The inset of (a) shows a detailed view of $\kappa$ around \emph{T}$_\texttt{N}$ and \emph{T}$_\texttt{S}$. The anisotropy of the correlation length $\eta_{\xi}$ as a function of temperature is shown in the inset of (b). 
 The dashed lines mark the locations of structural and magnetic transition temperatures \emph{T}$_\texttt{S}$ and \emph{T}$_\texttt{N}$.} 
\label{fig4} 
\end{figure}

      Below around \emph{T}$_\texttt{N}$, we notice that the charge correlation lengths along and normal to the in-plane component of AFM propagation wavevector (101)$_{\rm O}$ are different ($\xi_{a}>\xi_{b}$), displaying an anisotropic behavior. The anisotropy in charge correlation length 
is defined as\cite{Tucker2012}

\begin{equation}
\eta_{\xi}= \frac{\xi_a^2-\xi_b^2}{\xi_a^2+\xi_b^2}
\end{equation}
$\eta_{\xi}$=0 indicates isotropic correlations ($\xi_{a}$=$\xi_{b}$), whereas $\eta_{\xi}$=1 ($\eta_{\xi}$=-1) corresponds to the extreme case of $\xi_{a}\gg\xi_{b}$ ($\xi_{a}\ll\xi_{b}$) for structural domains consisting of 
long linear stripes. Based on this equation, we have derived the anisotropy of the charge correlation length as a function of temperature, as shown in the inset of Fig. 4. The anisotropy is most pronounced below \emph{T}$_\texttt{N}$, with values ranging from 0.1 to 0.8 as the temperature is lowered. 
           
        A possible scenario to interpret the anisotropic charge correlation lengths along $a$ and $b$ below around \emph{T}$_\texttt{N}$ in $x=0.047$ may be related to the presence of the antiphase magnetic domains \cite{Mazin2009,Yin2009,Li2012}.
Mazin and Johannes \cite{Mazin2009} first proposed that antiphase domains and their dynamic fluctuations are central for understanding the high-\emph{T}$_\texttt{C}$ ferropnictides. Very recently, Li \textit{et.al} \cite{Li2012} observed surface-pinnned antiphase domains in BaFe$_{2}$As$_{2}$  using high-resolution scanning tunneling microscopy.
 Since the energy differences between the AFM stripe magnetic structure and other AFM patterns are small \cite{Mazin2009}, it is highly possible that many antiphase magnetic boundaries are formed. 
The antiphase domains are pinned at \emph{T} $<$ \emph{T}$_\texttt{N}$, and show dynamic fluctuations in the region of \emph{T}$_\texttt{N}$ $<$ \emph{T} $<$ \emph{T}$_\texttt{S}$.
There are two kinds of simple antiphase domains (labeled A and B) with boundaries along $a$ and $b$ axes, as shown in Fig. 5. Due to the same magnetoelastic interactions that lead to a difference in the ferromagnetic and AFM bond lengths in the orthorhombic structure, we propose that the formation of such antiphase domains are accompanied by
elastic distortions at their boundaries. 
The antiphase boundaries along $a$ axis influence the magnitude of transverse charge correlation length $\xi_{b}$, whereas the antiphase boundaries along $b$ axis affect the longitudinal charge correlation length $\xi_{a}$. Differences in the density of antiphase boundaries in the two directions 
eventually leads to the anisotropy in $\xi_{a}$ and $\xi_{b}$ below \emph{T}$_\texttt{N}$. The dynamic fluctuations of the antiphase domains may be responsible for the anisotropy in charge correlation lengths
in the small temperature region above \emph{T}$_\texttt{N}$ (but lower than \emph{T}$_\texttt{S}$), as shown in the inset of Fig. 4 (b).

\begin{figure} \centering \includegraphics [width = 1\linewidth] {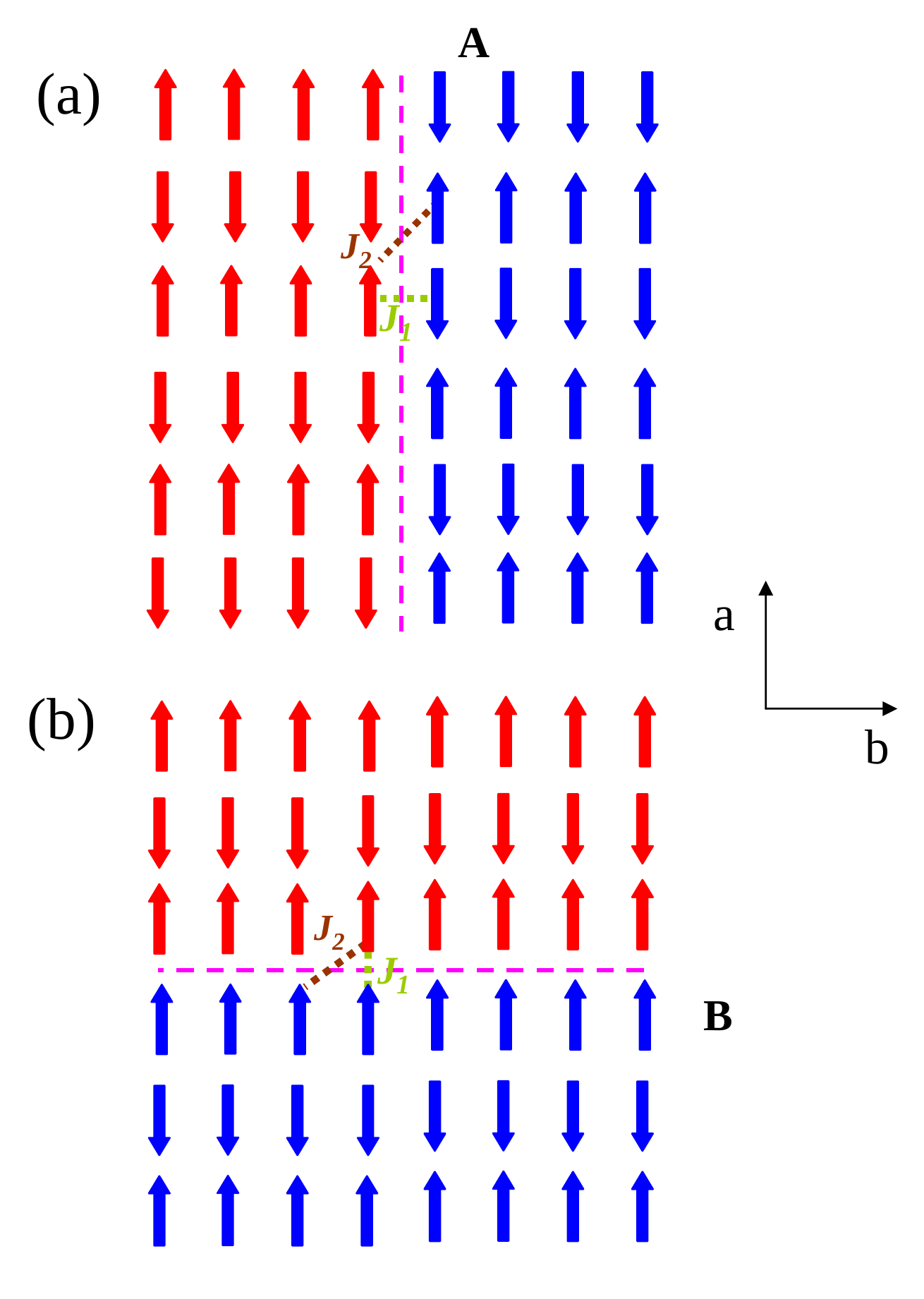}
\caption{(color online) Schematic pictures of two kinds of the antiphase magnetic domains, with boundaries along (a) $a$ axis and (b) $b$ axis, respectively. 
The antiphase domains are pinned at \emph{T} $<$ \emph{T}$_\texttt{N}$, but show the dynamic behavior in the region of \emph{T}$_\texttt{N}$ $<$ \emph{T} $<$ \emph{T}$_\texttt{S}$.  
The pink lines show the antiphase boundaries. \emph{J}$_\texttt{1}$ represents the nearest-neighbor exchange couplings along $a$ or $b$ directions, whereas 
 \emph{J}$_\texttt{2}$ represents the next-nearest-neighbor exchange couplings.}
\label{fig5} 
\end{figure}

      Based on such antiphase magnetic domain scenario and the $J_1-J_2$ model, the anisotropic charge correlation lengths can be used to estimate the ratio of the magnetic exchange parameters $J_1$ and $J_2$. As shown in Fig. 5, the energies per-spin (S) for forming these two kinds of antiphase boundaries are given by

\begin{equation}
E_{A}=(2 J_{2}-J_{1}) S^2
\end{equation}
\begin{equation}
E_{B}=(2 J_{2}+J_{1}) S^2
\end{equation}

 Since $J_1$ is AFM, the number of antiphase domain boundaries $N_{A}$ and $N_{B}$ should scale inversely with the magnetic energy of the domain wall so that 

\begin{equation}
\frac{N_{A}}{N_{B}}=\frac{E_{B}}{E_{A}}=\frac{2 J_{2}+J_{1}}{2 J_{2}-J_{1}}
\end{equation}
 The charge correlation length scales inversely proportional to the number of boundaries, i.e., 
$\xi_{a} \propto \frac{1}{N_{B}}$. Thus, 
     
\begin{equation}
\frac{\xi_{a}}{\xi_{b}}=\frac{N_{A}}{N_{B}}=\frac{2 J_{2}+J_{1}}{2 J_{2}-J_{1}}
\end{equation}

     In Fig. 4 (b), we observe $\xi_{a}/\xi_{b} \approx 3 $ at low temperatures and from Eq. (6) we can get $J_{2}\approx J_{1}$, which is consistent with the previous calculations or experiments on other iron pnictides, such as LaFeAsO \cite{Ma2008} and Ba(Fe$_{1-x}$Co$_{x}$)$_{2}$As$_{2}$ ($x=0.074$)\cite{Li2008}. This is also reasonable 
for producing a stripe-type AFM structure that requires  $J_{2}>\frac{J_{1}}{2}$. 

      It is worthwhile noting that the in-plane charge correlation lengths show gradual changes at temperatures significantly above \emph{T}$_\texttt{S}$, probably due to
 magnetic fluctuations that are known to persist above \emph{T}$_\texttt{S}$ in similar pnictides\cite{Diallo2010,Li2009}. We point out that this feature may support the nematic model.
In the nematic model\cite{Fernandes2010,Fernandes2012}, the nematic order coincides with the structural transition with the notion that the driving force 
for the  $T-O$ transition is not elastic in origin but magnetically driven by Ising-like interpenetrating AFM domains\cite{Fernandes2010,Paul2011}. Nematic (magnetic)
fluctuations remain at higher temperature above \emph{T}$_\texttt{S}$, which has been suggested by various techniques, such as susceptibility anisotropy \cite{Kasahara2012}, shear modulus\cite{Fernandes2010}, 
inelastic neutron scattering\cite{Harriger2011}, and anisotropic in-plane resistivity\cite{Chu2010}.

\begin{figure} \centering \includegraphics [width = 1\linewidth] {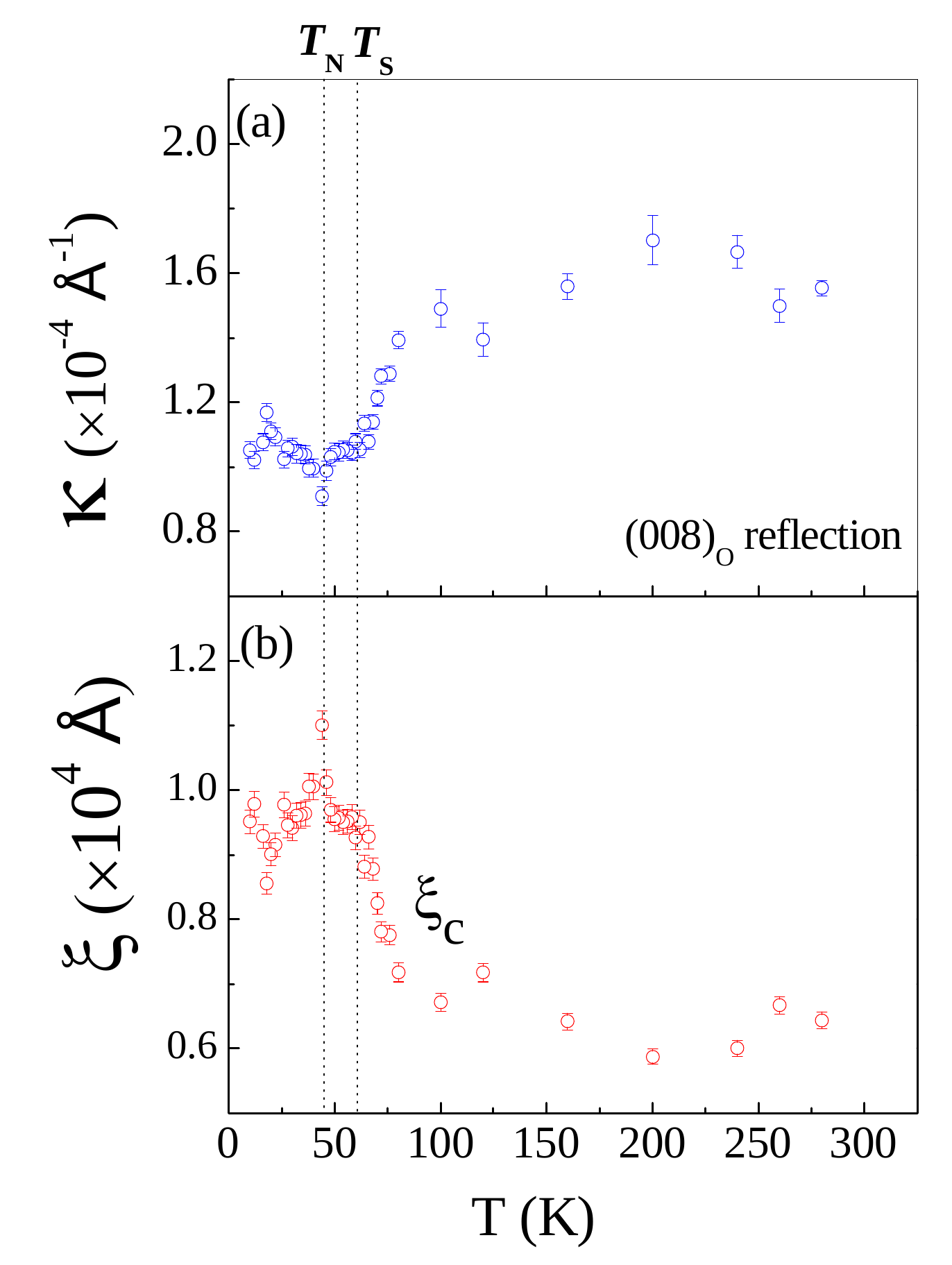}
\caption{(color online) Temperature dependence of (a) the intrinsic width $\kappa$ of the \emph{l} scan for (008)$_{\rm O}$
obtained from the Lorentzian function and (b) the out-of-plane  charge correlation length along $c$ axis ($\xi_{c}$).The dashed line indicates the locations of \emph{T}$_\texttt{N}$ and \emph{T}$_\texttt{S}$.}
\label{fig6} 
\end{figure}  

   We now turn to a discussion of the $c$-axis charge correlation length. Fig. 6(a) and (b) show the temperature dependence of the intrinsic width of the (008)$_{\rm O}$ Bragg reflection and the corresponding out-of-plane  charge correlation length along $c$ axis  ($\xi_{c}$), respectively. A single anomaly 
at the magnetic transition temperature \emph{T}$_\texttt{N}$ and a gradual change above and through \emph{T}$_\texttt{S}$, are observed. The absence of sharp anomaly in $\xi_{c}$ suggests that atomic distortions resulting from the T-O structural
 transition  mainly occur in the $ab$ plane. This is consistent with the fact that the in-plane lattice parameters $a$ and $b$ change significantly, but $c$ changes weakly around \emph{T}$_\texttt{S}$, as can also be seen from Fig.\ \ref{fig2}.
Both the out-of-plane charge correlation length $\xi_{c}$ and lattice parameter $c$ show an anomaly at \emph{T}$_\texttt{N}$, showing close correlation between
the AFM magnetic transition and the modification of structure along $c$ axis. Recent experiments and calculations reveal that the Fe local moment is very sensitive to the Fe-As 
distance in iron pnictides. Yin \emph{et al.} \cite{Yin2008} performed density functional theory (DFT) calculations within the generalized gradient approximation and found the Fe-Fe transverse exchange coupling is strongly
dependent on both the AFM symmetry and the Fe-As distance. Belashchenko \emph{et al.}\cite{Belashchenko2008} demonstrated that in layered iron-pnictide compounds, as the Fe-As distance is decreased, 
the degree of itinerancy of Fe moments increase. Moreover, the coupling between the local moment and the Fe-As distance is controlled by strong covalent Fe-As bonding. Recently,
 neutron diffraction studies of CeFeAs$_{1-x}$P$_{x}$O \cite{Cruz2010} and DFT calculations \cite{Lee2010} demonstrated that a decrease in Fe-As distance induces strong 
hybridization between  Fe 3\textit{d} and As 4\textit{p} orbitals, leading to quenched Fe magnetic moments. Therefore, the AFM transition at \emph{T}$_\texttt{N}$ is coupled to the change of
 Fe-As distance, which leads to the anomalies in the out-of-plane charge correlation length and lattice parameter $c$.

      In summary, high resolution X-ray diffraction studies on structural Bragg reflections of the SC and AFM Ba(Fe$_{1-x}$Co$_{x}$)$_{2}$As$_{2}$ ($x$=0.047) single crystal reveal secondary effect
stemming from the magnetic properties of the system, which is understood to result from intrinsic and strong magnetoelastic coupling. In addition to showing anomalies around the structural and magnetic transitions, the in-plane charge
 charge correlation lengths along $a$ and $b$ axes show anisotropy below around \emph{T}$_\texttt{N}$, which probably results from the effect of antiphase boundaries formed along $a$ and $b$ axes. Employing our anisotropic charge correlation lengths, we are able
 to estimate the ratio of $J_{2}/ J_{1}$ to be around 1 on the basis of such antiphase magnetic domain scenario and $J_1-J_2$ model. The out-of-plane charge correlation length $\xi_{c}$ and lattice parameter $c$ exhibit a single anomaly at \emph{T}$_\texttt{N}$, which can be associated with the modification of Fe-As distance when the AFM transition occurs. 
Our results also show gradual evolution of the Bragg peak widths and in-plane charge correlation length above \emph{T}$_\texttt{S}$, which is presumably induced by the nematic magetic fluctuations up to almost 200 K. 

\section{Acknowledgements}

Research at Ames Laboratory is supported by the US Department of Energy, Office of Basic Energy Sciences, Division
of Materials Sciences and Engineering under Contract No. DE-AC02-07CH11358. Use of the Advanced Photon Source
at Argonne National Laboratory was supported by the US Department of Energy, Office of Science, Office of Basic
Energy Sciences, under Contract No. DE-AC02-06CH11357

\end{document}